## Highlights

## First-Principle Analysis of Hydrogen Storage Potentials in Vanadium Hydride Perovskites $XVH_3$ (X = Li, K)

Anupam, Shyam Lal Gupta, Sumit Kumar, Ashwani Kumar, Sanjay Panwar, Diwaker

- Stability of $XVH_3$ (X =Li,K ) hydride perovskites is confirmed by Enthalpy of formation.
- The gravimetric hydrogen storage capacity for $XVH_3$ (X =Li,K ) are 3.25% and 4.97%
- Metallic behavior is revealed by electronic structure of these hydrides.
- Enhanced potential for alternative green energy source.

# First-Principle Analysis of Hydrogen Storage Potentials in Vanadium Hydride Perovskites $XVH_3$ (X = Li, K)

Anupam[a], Shyam Lal Gupta[b], Sumit Kumar[c], Ashwani Kumar[d], Sanjay Panwar[a,*] and Diwaker[e,**]

[a]*School of Basic and Applied Sciences, Maharaja Agrasen University, Baddi, Solan, 174103, H P, INDIA*
[b]*Department of Physics, HarishChandra Research Institute, Prayagraj, Allahabad, 211019, U P, INDIA*
[c]*Department of Physics, Government College, Una, 174303, H P, INDIA*
[d]*Department of Physics, Abhilashi University, Mandi, 175045, H P, INDIA*
[e]*Department of Physics, SCVB Government College, Palampur, Kangra, 176061, H P, INDIA*

ARTICLE INFO

ABSTRACT

This study investigates Vanadium-based $XVH_3$ (X = Li,K) hydride perovskites for their hydrogen storage capacity using the WIEN2K code. Structural studies reveal that these hydrides are stable and belongs to cubic space group (221 Pm-3m). We have examined many aspects of these compositions throughout, using the PBE-GGA and mBJ exchange correlation potential. The study also examines their thermodynamic stability and gravimetric hydrogen storage capacity. The findings suggest metallic nature of these compositions and strong thermoelectric responses, making them potential candidate for green energy sources.

## 1. Introduction

Perovskites with exceptional qualities have become one of the most fascinating new materials for the twenty-first century. Perovskite has attracted a lot of attention and made substantial strides in energy storage, pollutant degradation, and optoelectronic devices in the last several decades because to its remarkable photoelectric and catalytic properties[1]. Materials of the empirical formula $ABX_3$ are referred to as perovskites. These materials are further divided into two groups: organic-inorganic hybrid perovskite and inorganic perovskite. $SrTiO_3$ is a prime example of a perovskite due to its unique structure in the lower symmetric space group. The recent studies on various transition metal hydrides, complex hydrides and carbon based compounds can be act as better alternative for green energy demand globally [2, 3, 4, 5, 6, 7, 8, 9, 10, 11, 12, 13, 14, 15]. Perovskite, with its characteristic magnetic and electric properties, holds potential for developing novel hydrogen storage applications due to its affordable cost and fascinating physical and chemical features. Alkali and alkaline earth metals, A and B, respectively, make up perovskites hydrides, which are ionic compounds with predicted band gaps of at least 2 eV. The stability trend of tolerance factors largely supports Goldschmidt's theory. The chemical compound known as perovskite hydrides, or $ABX_3$, has drawn interest lately as a potential viable hydrogen storage solution. Although perovskite materials have significant role in storing Hydrogen but no such significant compounds found in study or literature yet. So the materials which met the Hydrogen storing requirements are still being continuously searched [16, 17, 18, 19, 20, 21, 22, 23, 24, 25, 26, 27, 28, 29, 30]. Materials for storing hydrogen often have lot of hydrogen bonding in these materials. They have four key characteristics: (1) they are large enough to hold large amounts of hydrogen; (2) they possess catalytic qualities that improve hydrogen absorption; and (3) they have adequate gravimetric hydrogen storage capacities that improve hydrogen storage. The gravimetric densities of hydroxide perovskites typically range from 1.2 to 6.0 weight percent. Hydrogen may be stored more effectively and safely in metal hydride perovskites than in liquid or compressed gas phase [31, 32, 33, 34, 35, 36]. This study explores the structural, electronic, thermodynamic, thermoelectric, mechanical, and hydrogen storage characteristics of perovskite hydrides with a specific pairing $XVH_3$ (X = Li, K) for hydrogen storage applications using density functional theory.

## 2. Computational details

The $XVH_3$ (X = Li, K) cubic phase structure is taken into consideration for all compositions by using space group 221 (Pm-3m) to investigate different properties using WIEN2k [37, 38]. The structural properties were calculated using Perdew Burke Ernzerhof (PBE) and the generalized gradient approximation (GGA) while the electronic behaviour is explored using mBJ functional. This is beacause modified Becke-Johnson (mBJ) functional is reported to predict reliable band structure [39] even more correctly and fast compared to the usual hybrid functional approach [40, 41, 42] in Wien2K. Birch-Murnaghan's equation of state [43, 44, 45, 46] is used to explore the structural characteristics and determine the stable phase of these perovskites. All these calculations uses 216 k-points, an RMT of 5.0, and an energy cutoff of -6.0 Ryd. Geometrically optimized calculations were used to compute the structural, electronic, thermodynamic, thermoelectric, mechanical, and hydrogen storage properties of cubic compositions $XVH_3$ (X = Li, K).The WIEN2k code has been used for all calculations.

*Corresponding author
**Principal corresponding author
dr.spanwar@gmail.com (S. Panwar); diwakerphysics@gmail.com ( Diwaker)
ORCID(s): 0000-0003-3283-3165 (S. Panwar); 0000-0002-4155-7417 ( Diwaker)

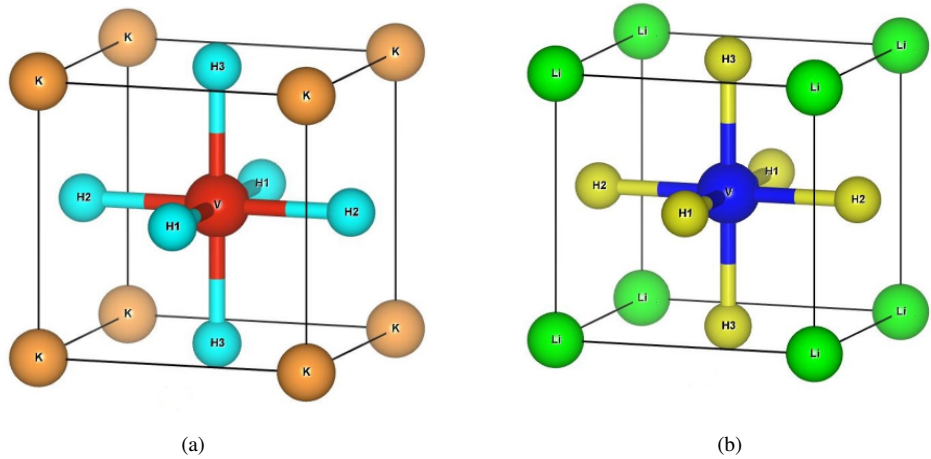

Figure 1: Optimised crystal structure of (a) KVH$_3$ (b) LiVH$_3$

**Table 1**
Lattice constants [a (Å)], Bulk Modulus ( B in GPa), Volume (V$_0$ in Å$^3$), Pressure derivative of Bulk modulus ( B$'$ in GPa), Ground state Energy (E$_0$ (eV)), enthalpy of formation ($\Delta H_f$ (KJ/mol.H$_2$)), decomposition temperature (T$_d$) and Gravimetric storage densities (C$_{wt\%}$) of XVH$_3$ with (X=Li, K) perovskites.

| Perovskite | a | B | V$_0$ | B$'$ | E$_0$ | $\Delta H_f$ | T$_d$ | C$_{wt\%}$ |
|---|---|---|---|---|---|---|---|---|
| KVH$_3$ | 3.8341 | 56.99 | 380.36 | 3.97 | -3106.31 | -457.27 | 3499 | 3.25 |
| LiVH$_3$ | 3.4505 | 80.15 | 277.21 | 4.62 | -1917.18 | -248.86 | 1904 | 4.97 |

Thermal investigations has been done using BoltzTrap2 code[47] while Gibbs 2 code[48] is used for thermodynamic investigations.

## 3. Results and Discussions

This section will explore the structural, electronic, thermodynamic, thermoelectric, mechanical, and hydrogen storage properties of vanadium-based XVH$_3$ hydride perovskites.

### 3.1. Investigations of Structural Phase Stability on XVH$_3$ (X = Li, K)

The proposed hydride perovskites adopt FCC symmetry with space group 221(Pm-3m), with a general formula ABH$_3$, where A is a cation usually Alkali metals (Li and K in our case) while B is a transition metal (V in our case) which forms a octahedral coordination with different hydrogen atoms. The Wyckoff positions of A species (Li and K in our case) occupy Wyckoff positions 1a (0,0,0) and B species which is V in our case is placed at Wyckoff positions 1b (0.5,0.5,0.5) while the three hydrogens are at Wyckoff positions 3c as ( (0, 0.5, 0.5), (0.5, 0, 0.5), and (0.5, 0.5, 0), respectively. Figure 1 shows the relaxed crystal structure of XVH$_3$ (X = Li, K) hydrides perovskites. Figure 2 (a-b) displays energy vs volume graphs that illustrate the minimal ground state energy of these compositions. Table 1 lists parameters like Lattice constants, Bulk Modulus, Volume, Pressure derivative, Ground state Energy, enthalpy of formation, decomposition temperature, and Gravimetric storage densities of XVH$_3$ (X= Li, K ) perovskites.

### 3.2. Stability, enthalpy, decomposition temperature and hydrogen storage of the inter-metallic hydrides

The stability of intermetallic hydrides has been predicted using the enthalpy of production ($\Delta H_f$) and the decomposition temperature (T$_d$). Phase stability and dehydrogenation properties are found for XVH$_3$ (X = Li, K) hydrides perovskites. The reactions that result in XVH$_3$ (X = Li, K) perovskites may be comprehended using the following equations.

$$KVH_3 \leftrightarrow KV(s) + \tfrac{3}{2} * H_2(g) \qquad (1)$$

$$LiVH_3 \leftrightarrow LiV(s) + \tfrac{3}{2} * H_2(g) \qquad (2)$$

Using Hess law as given below

$$\Delta H_f = \sum E_{tot}(\text{products}) - \sum E_{tot}(\text{reactants}) \qquad (3)$$

The enthalpy of formation for these compositions with (X=Li,K) is calculated using the equation given below

$$\Delta H_f = E_{tot}(XVH_3) - E_{tot}(XV) - \tfrac{3}{2} E_{tot}(H_2) \qquad (4)$$

Notably, we get a formation enthalpy of -457.27kJ/mol.H$_2$ and -248.86 kJ/mol.H$_2$ using the FP-LAPW technique for (X=K, Li) proving these perovskites' exceptional thermodynamic stable. Further, the standard Gibb's energy ($\Delta$G) relation is given as

$$\Delta G = \Delta H_f - T\Delta S \qquad (5)$$

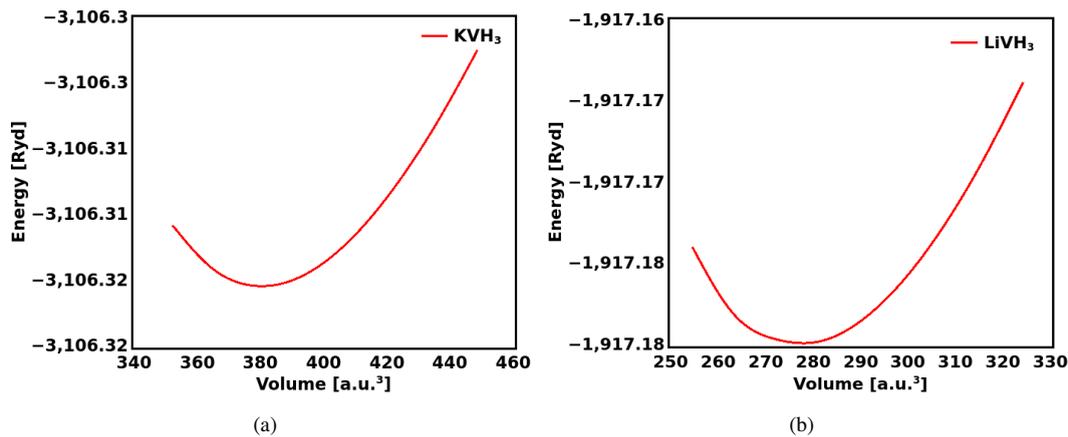

**Figure 2:** Energy versus Volume curves of (a) KVH$_3$ (b) LiVH$_3$

where $\Delta H_f$ is fomation enthalpy and $\Delta S$ is standard entropy change for hydrogen. At equilibrium, the standard value of Gibb's energy is zero. Further, if we talk about the value of change in entropy i.e $\Delta S$ it results from the way the hydrogen molecule changes during the dehydrogenation reaction as $\Delta S = 130.68 J/mol.K^{-1}$ [49]. Therefore, the thermal decomposition $T_d$[49] can be calculated using the relation given as

$$T_d = -\frac{\Delta H_f (in J/mol)}{130.68} \quad (6)$$

The calculated decomposition temperature for KVH$_3$ was 3499 K and for LiVH$_3$ was 1904 K using the enthalpy of formation value from Table 1 in the aforementioned equation. These indicate the potential of studied compositions for high temperature hydrogen cyclability. Understanding the gravimetric hydrogen storage capacity of these compositions is crucial for their application in hydrogen storage applications. The term "gravimetric hydrogen storage capacity" refers to the amount of hydrogen that may be stored per mass unit of a substance. Using the equation below, the gravimetric hydrogen storage capacities $C_{wt\%}$[50] for the compositions under study are determined.

$$C_{wt\%} = \left[\frac{n_H * M_H}{M_{(XVH_3)}} \times 100\right] \% \quad (7)$$

In this case, the number of hydrogen atoms is $n_H$, the molar mass of hydrogen is $M_H$, and the molar mass of the compound is $M_{(XVH_3)}$. From the above formula after calculation, the gravimetric hydrogen storage capacity is found to be 3.25% and 4.97% for XVH$_3$ (X=Li, K) respectively. . These are close to the DOE set targets of 5.5 % till end of 2025 for light duty fuel cell vehicles as reported by US Department of energy. The authors have investigated one formula unit of our compositions which shows these gravimetric hydrogen capacity but if we consider thin films of these materials they have enhanced gravimetric hydrogen capacity [31, 33, 32].

### 3.3. Energy band topologies, density of state plots

The amazing electronic properties of perovskites, a class of materials with a distinctive crystal structure, have garnered a lot of interest lately. Perovskite materials, due to their tunable bandgap, are versatile and can be utilized in various applications such as photodetectors, solar cells, LEDs, and hydrogen storage. Along the high symmetry sites in the first Brillouin zone, the energy band structure for XVH$_3$ (X = K, Li) hydride perovskites is computed and displayed in figures 3(a-b). In order to compare the strength of the electronic correlations among different species in a particular composition, we also calculated the band structure using PBE-GGA and compared with those obtained from mBJ approach (See appendix 1). It can be seen from figures as reported in appendix 1 that both approaches lead to the same band structure. This indicates that these are weakly correlated systems. D.O.S. is a fundamental concept in solid-state physics, indicating the density of electronic states in a material at a specific energy level. DFT investigation reveals metallic compounds with no energy gap between valance and conduction bands. Total (T.D.O.S.) and partial density of states (P.D.O.S.) of XVH$_3$ were calculated for electronic contribution. Figure 4 (a-c) shows the plots T.D.O.S. and P.D.O.S. of XVH3 (X = K, Li) hydride perovskites. For KVH$_3$ and LiVH$_3$, the greatest value of T.D.O.S. in the valence band is 0.9 states/eV and 0.75 states/eV, respectively. The understanding of the electronic structure of materials can be facilitated by the partial density of states. It offers details on the density of states of a particular element or material's orbital. We note that a significant contribution has come from the d-states in KVH$_3$ and p states in LiVH$_3$. This can be attributed to different electropositivity of different alkali metals present in these compositions.

### 3.4. Lattice dynamic stability

The dynamical stability of metal hydride XVH$_3$ with (X=Li, K) has been determined via phonon dispersions in the Brillouin zone as represented in Figure. 5(a-b). The optical behaviour of hydride materials is caused by the optical branches, which are found in the top region of the dispersion curves. These optical modes are the result of atoms in a lattice oscillating out of phase. Conversely, acoustic branches, which are located at the bottom of phonon dispersion curves, originate from the coordinated vibration of atoms in a lattice beyond their balance position. According to literature stable structure only contains positive vibration states within phonon dispersion curves. The crystal bonding energy will be lost in the negative vibration states, and occasionally the structure will vanish as a result of breaking of atomic bonds. Therefore, the absence of negative frequencies throughout the whole Brillouin zone of Fig. 5(a-b) in the phonon dispersion curves strongly implies that XVH$_3$ with (X=Li, K) hydrides dynamically stable [51].

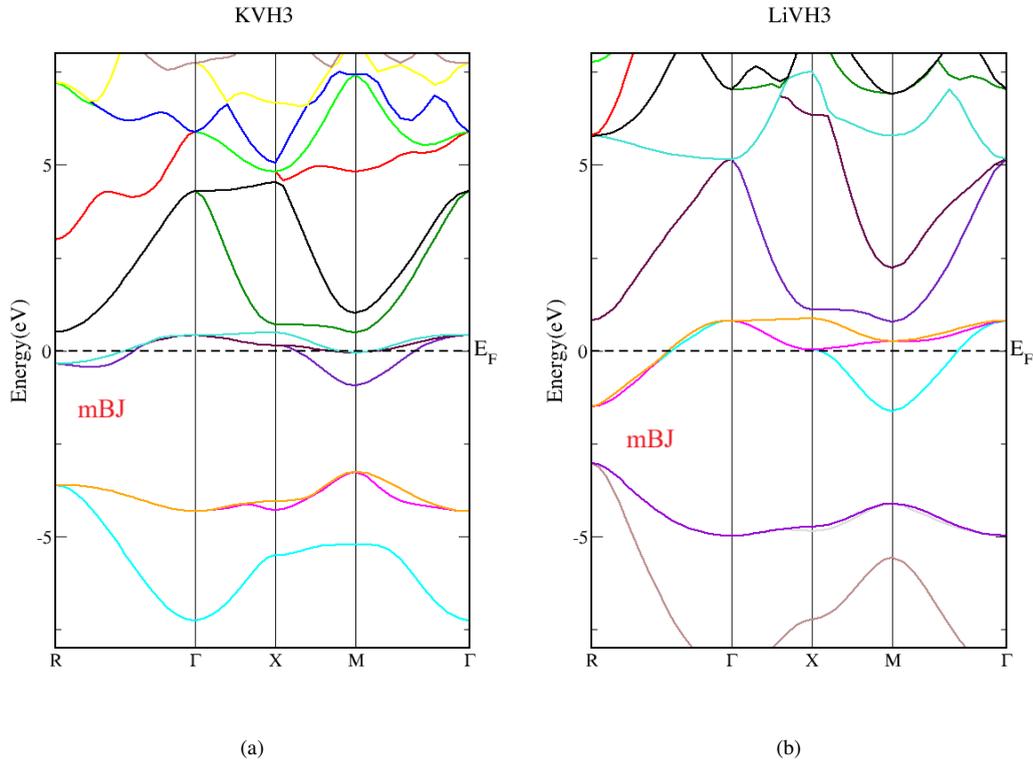

**Figure 3**: Energy Band structures using mBJ approach [a] KVH$_3$ [b] LiVH$_3$

### 3.5. Thermoelectric Properties

The world is now dealing with an energy crisis. Thermoelectric materials, which convert waste heat into electrical energy, are reported as eco-friendly, intelligent, and clean solutions to address the energy crisis[1]. Different thermoelectric properties are calculated as a function of $(\mu - \epsilon_F)$ at different temperatures, including the Seebeck coefficient (S), electrical conductivity $\sigma$, electronic thermal conductivity ($\kappa_e$), power factor ($\frac{S^2\sigma}{\tau}$) [1]. The outcomes, which guarantee our compositions' potential for such applications, are displayed in Fig. 6 and 7. When a temperature gradient is introduced between two junctions, a phenomenon known as see beck coefficients is produced, which results in voltage. The Seebeck coefficient is displayed in Fig. 6a and 7a for our compositions XVH$_3$ (X = K, Li). Plots show that at temperatures of 700 K and 300 K, respectively, the highest value of the Seebeck coefficient is S~.0015 VK$^{-1}$ for both KVH$_3$ and LiVH$_3$ and no significant change is observed in its value with variation in temperature. The electrical conductivity variation for XVH$_3$ (X = K, Li) as a function of chemical potential is shown in Fig. 6b and 7b. The materials' thermoelectric qualities are closely correlated with their electrical conductivity. Electrical conductivity, or the passage of electrons from high-temperature areas to low-temperature areas, is what produces current. At a temperature of 100K, the highest values of $\sigma$ ~1.3 × 1e21 $\Omega^{-1}m^{-1}s^{-1}$ and 1.5 × 1e21 $\Omega^{-1}m^{-1}s^{-1}$, respectively, are found for these compositions, XVH$_3$ (X = K, Li). There are no appreciable variations in conductivity with rise in temperature. Furthermore, as shown by Fig. 6b and 7b, the equivalent value of electrical conductivity is also changing as the chemical potential increases. Thermal conductivity is the ability of a material to transfer heat which is a result of electron mobility and lattice vibrations. Fig. 6c and 7c, which fluctuate according to chemical potential at various temperature ranges, depict the thermal conductivity. The thermal conductivity peak values for XVH$_3$ (X = K, Li) at 900 K were around 1.2 × 10$^{16}$ $Wm^{-1}K^{-1}s^{-1}$ and 1075 × 10$^{16}$ $Wm^{-1}K^{-1}s^{-1}$, respectively. The correlation between temperature and thermal conductivity makes perovskite ideal for high temperature thermoelectric applications due to their close compatibility. The partial efficiency of a thermoelectric device is commonly determined by its power factor (PF). Figures 6d and 7d display the (PF) for compositions XVH$_3$(X = K, Li) as a function of chemical potential at different temperature ranges. The PF peaks at 100 K as the temperature rises. We deduced from the plots that the best values of (PF) for XVH$_3$ (X = K, Li) at temperature 100K were around $5 \times 10^{12} Wm^{-1}K^{-2}s^{-1}$ and $8 \times 10^{12} Wm^{-1}K^{-2}s^{-1}$ respectively. The study presents a theoretical approach for hydrogen storage using thermoelectric compositions, revealing their strong thermoelectric responses and potential for green energy sources.

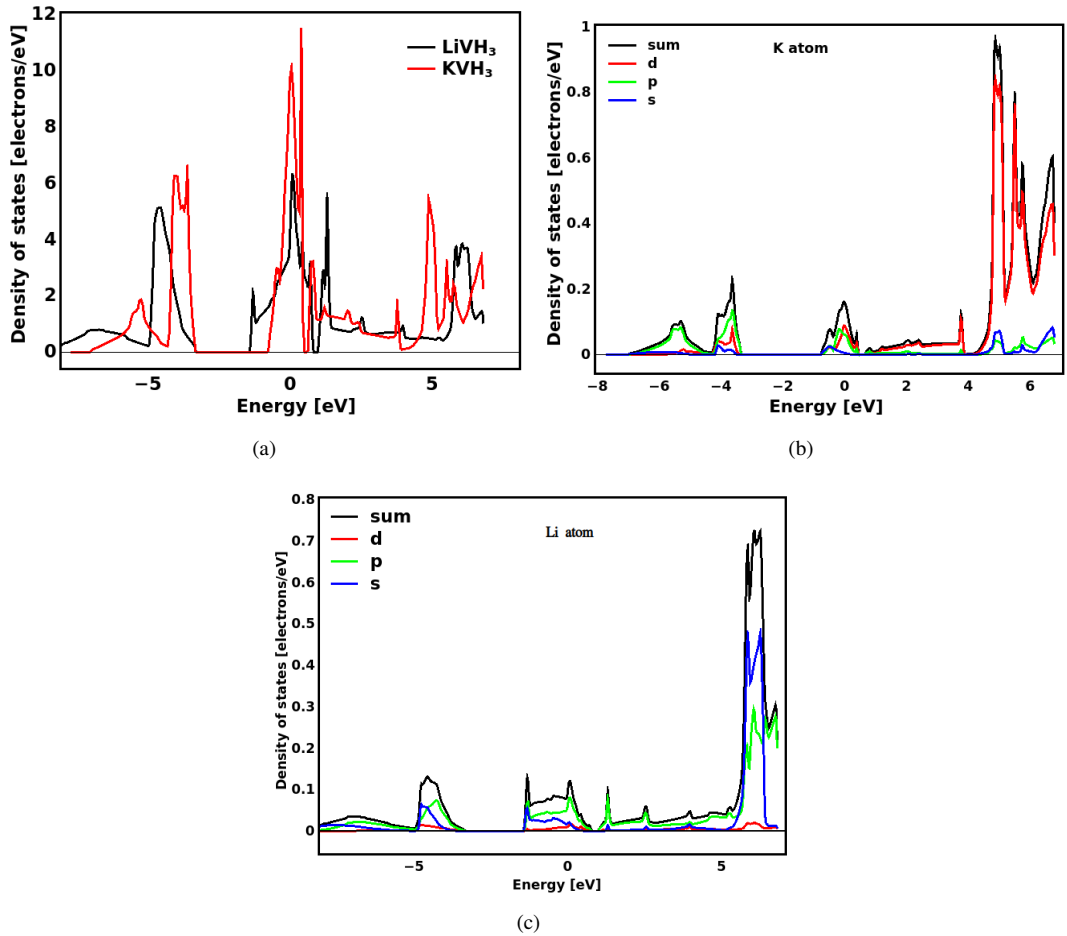

**Figure 4**: [DOS/PDOS for XVH$_3$ (X=K, Li) [a] TDOS [b] K atom PDOS [c] Li atom PDOS

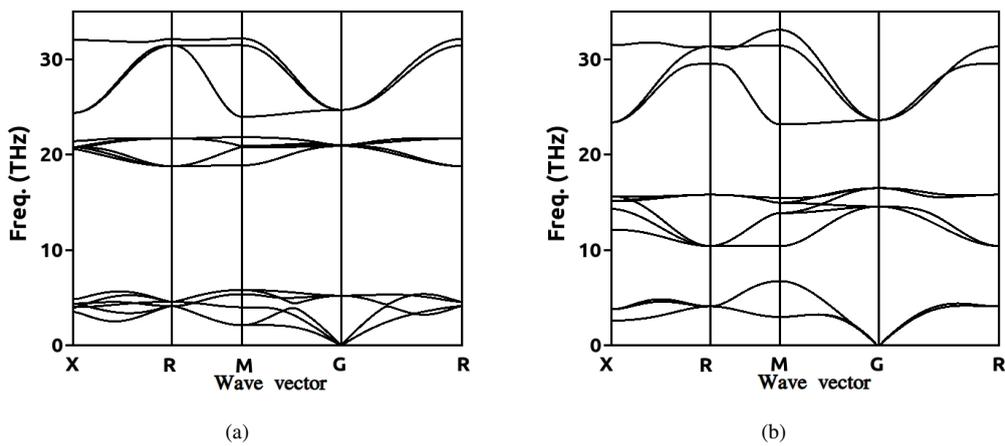

**Figure 5**: The phonon dispersion curves for [a] KVH$_3$ [b] LiVH$_3$

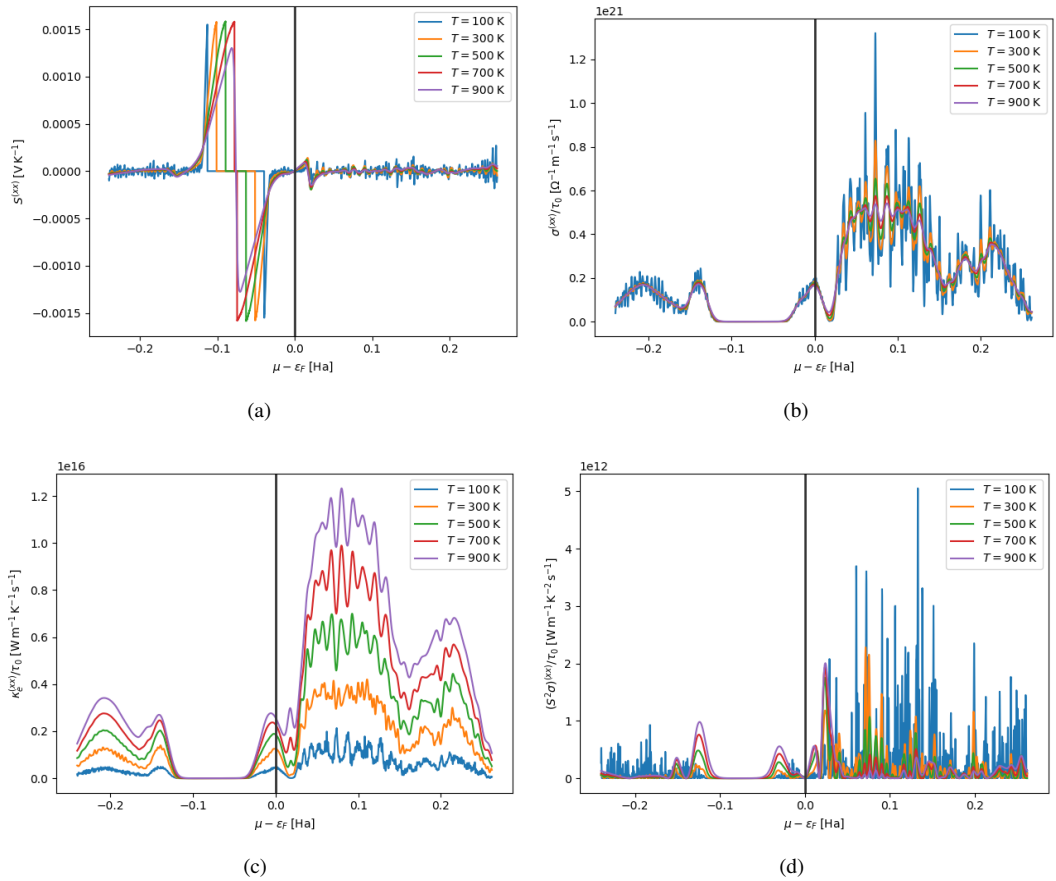

**Figure 6**: [a] Seebeck coefficient (S), [b] Electrical Conductivity ($\sigma$), [c] Electronic thermal conductivity ($\kappa_e$), [d] Power Factor ($S^2\sigma$) as a function of ($\mu - \epsilon_F$) at different temperatures for KVH$_3$

### 3.6. Optical Properties

In general, the material having nonzero electronic DOS and the non-zero band gap at the fermi level is known to show the metallic character and is characterized via the electronic dynamics measurements (energy transport) along the field direction under some applied bias through a parameter called as longitudinal conductivity. On the other hand the optical conductivity originates from the interaction of material with electromagnetic radiations which have the field in transverse direction to the energy transportation. Thus, there is always possibility of getting results different than the intuitive ones. This imputed us to explore the dielectric and optical properties even after we have observed the metallic character in electronic band structure and DOS of studied compositions. These interesting and intriguing observations could be helpful for researchers exploring materials for specific frequency dependent applications such as wavefront engineering, electromagnetic tunnelling, electromagnetic cloaking etc. This section primarily focuses on the optical characteristics of materials to comprehend their interaction with electromagnetic photons [52]. The study aims to evaluate the suitability of hydroxide perovskite materials for optoelectronic applications, particularly hydrogen storage, by analyzing optical parameters based on radiation frequency. The absorption coefficient, $\alpha(\omega)$, is a crucial factor that determines the distance light can travel through a material, influenced by the incoming light's wavelength. High absorption coefficients of materials absorb light photons more quickly, thereby stimulating the electrons in their conduction band. Fig. 8 (a) displays our calculated energy-dependent absorption coefficients for hydride perovskites, namely XVH$_3$ (X = Li, K). Remarkably, a notable rise in the absorption rate of LiVH$_3$ is observed with increasing incident radiation frequency. The maximum absorption coefficient values for XVH$_3$ (X = Li, K), near the infrared region are 2x10$^5$ and 1.7x10$^5$ respectively, based on the presented graphs. A material surfaces ability to reflect incoming radiation is measured by a property called reflectivity. We investigated the frequency-dependent reflectivity R($\omega$) of hydride perovskites and plotted graphs are shown in Figure 8(b). The peak reflectivity values for XVH$_3$ (X = Li, K) are 0.65 and 0.45 in the 12 to 14 eV range. The reflectance decreased rapidly as the frequency increased in the high-frequency band. The amount of energy lost by scattering or dispersion during an electron's transition is expressed by the energy loss function L($\omega$). The correlation is based on the scattering probabilities occurring during inner shell transitions. The optical loss function of hydride perovskites, XVH$_3$, is plotted against the incident

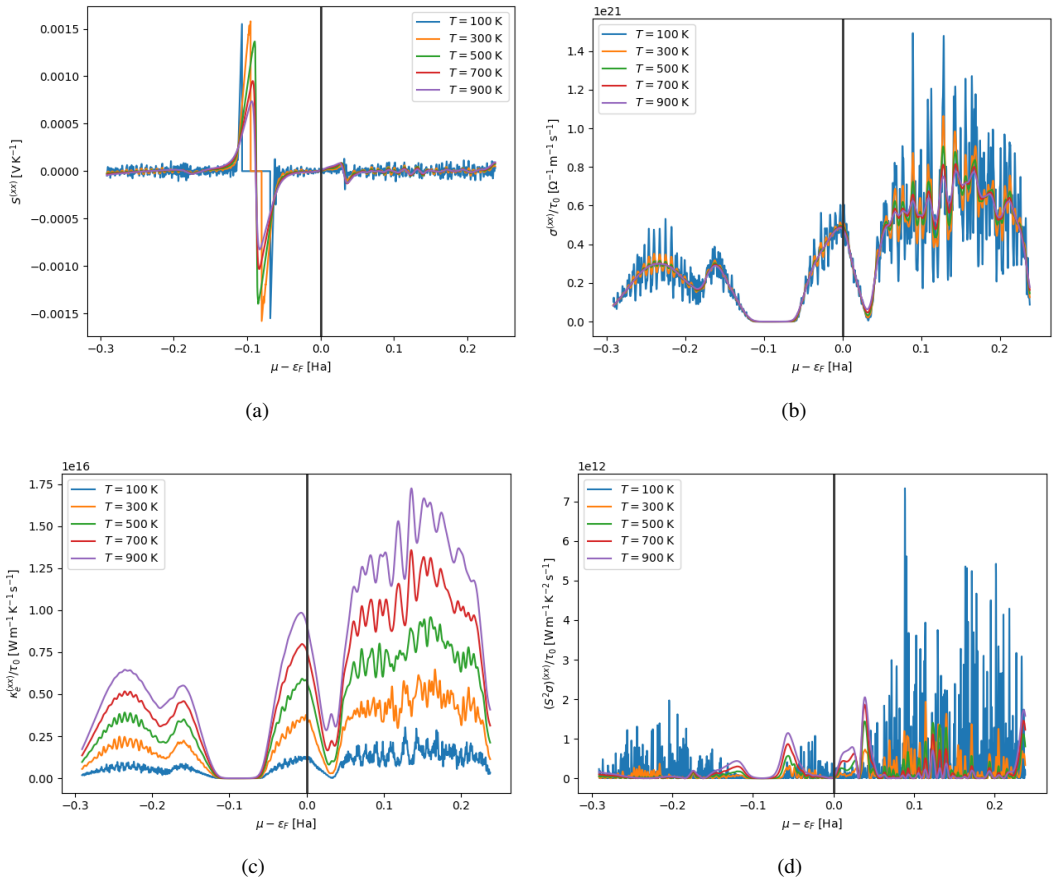

**Figure 7**: [a] Seebeck coefficient (S), [b] Electrical Conductivity ($\sigma$), [c] Electronic thermal conductivity ($\kappa_e$), [d] Power Factor ($S^2\sigma$) as a function of ($\mu - \epsilon_F$) at different temperatures for LiVH$_3$

photon energy in Fig. 8(c). This plot exhibits a peak optical loss values of 0.9 and 1.5 at a photon energy of 2.5 and 1.9 eV for XVH$_3$ (X = Li, K). Plasmon emission is most likely to blame for the abrupt drop in Loss function[52] that was seen at high frequencies for all of the hydride perovskites compositions under study. It is also important to note that LiVH$_3$ has lesser values of loss function at high photon energy compared to KVH$_3$. Hence, LiVH$_3$ could be potential candidate for high frequency optical applications. A recent study employed optical conductivity $\sigma(\omega)$ to investigate how incident radiation interacts with the surface of XVH$_3$ (X = Li, K)hydride perovskites to break bonds and cause conduction through the photoelectric effect. The conductivity of hydride perovskites XVH$_3$ (X = Li, K) as a function of frequency is plotted against the incident photon energy, which ranges from 0 to 14 eV, as shown in Figure 8(d). LiVH$_3$ demonstrated the highest conductivity value, reaching 27 (1/fs) at a photon energy of 0.2 eV. The optical examination of the investigated compounds indicates that LiVH$_3$ is a promising material for hydrogen storage due to its high photon conductivity at low energy.

### 3.7. Mechanical Properties

To explore the characteristics of these compositions under high hydrostatic pressure we have calculated and reported the elastic constants as obtained using the model implemented in IRelast package integrated with Wien2k[53] for compressive force (pressure) of 5GPa and 10 GPa. These constants are calculated at a given pressure by computing the stress tensor components for tiny deformation and applying energy, in line with a lattice deformation that maintained volume [54] to provide useful information about a compound's toughness and stability. These could be experimentally obtained using Diamond Anvil Cell [55]. The strain-dependent matrix of second-order elastic constants ($C_{ij}$), equilibrium volume, and crystal energy are some of the variables that affect a lattice's elastic behavior. A symmetric 6X6 stiffness matrix with 21 independent components represents the material's response to applied stresses in accordance with Hooke's law. Table 2 provides information on the key elastic constants for the cubic perovskites XVH$_3$ (X = K, Li) i. e. $C_{11}$, $C_{22}$, and $C_{44}$ [56]. The mechanical and elastic properties of individual perovskites compositions are largely determined by these components. The Born stability criteria, commonly known as the mechanical stability conditions, must be met by a compound having a mechanically and elastically stable structure. These criteria are $C_{11} > 0$, $C_{44} > 0$, $(C_{11}-C_{12})>0$, $(C_{11}+2C_{12})>0$ [57]. Table 2 presents three unique constants that meet the Born stability requirements, signifying the mechanical stability of every perovskite compound under investigation. Cauchy's pressure is calculated by the following equation given as

$$CP = (c_{12} - c_{44}) \quad (8)$$

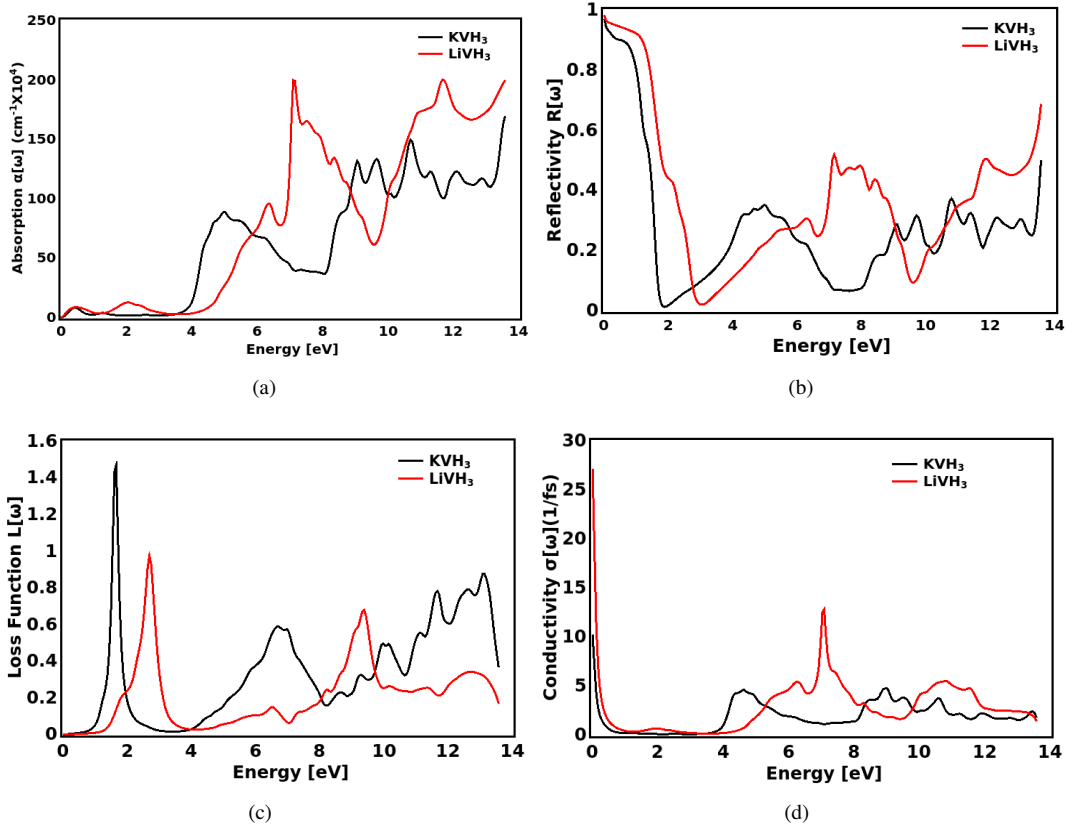

**Figure 8:** optical properties for XVH$_3$(X=Li,K) perovskite type hydrides [a] Absorption, [b] Reflectivity, [c] Loss function, [d] Conductivity

B is computed as by using the V.R.H. technique:

$$B = \frac{1}{3}\left[c_{11} + 2c_{12}\right] = \frac{1}{3}\frac{1}{(s_{11} + 2s_{12})} = B_V = B_R \quad (9)$$

where

$$s_{11} = \frac{c_{11} + c_{12}}{c}; s_{12} = \frac{-c_{12}}{c}; \quad (10)$$

and

$$c = (c_{11} - c_{12})(c_{11} + 2c_{12}); s_{44} = \frac{1}{c_{44}} \quad (11)$$

Also

$$B = \frac{B_v + B_R}{2} \quad (12)$$

The following relation is used to determine Young's relation as

$$E = \frac{9GB}{3B + G} \quad (13)$$

Shear modulus is computed using the following relation as

$$G = \frac{G_R + G_v}{2} \quad (14)$$

where $G_R$ and $G_v$ are

$$G_V = \frac{1}{5}(c_{11} - c_{12} + 3c_{44}) \quad (15)$$

and

$$G_R = 5\left[\frac{(c_{11} - c_{12})c_{44}}{3(c_{11} - c_{12}) + 4c_{44}}\right] \quad (16)$$

Poisson's ratio ($v$) are determined using the elastic constants given by

$$k = \frac{B}{G}; G = \frac{c_{11} - c_{12} + 3c_{44}}{5} \quad (17)$$

and

$$v = \frac{c_{12}}{c_{11} + c_{12}} \quad (18)$$

The shear modulus (G) and Poisson's ratio ($v$) are used to calculate the Young's modulus given as

$$E = 2G(1 + v) \quad (19)$$

**Table 2**
Elastic constants (C$_{ij}$), Bulk, Shear and Young Modulus (B$_V$, B$_R$, B, G$_V$, G$_R$, G and E$_V$, E$_R$, E in GPa), Reuss and Hill Poisson's coefficient($v_V$, $v_R$ in GPa), Kleinman's parameter ($\zeta$), Ranganathan and Kube Anisotropy Index ($A_u$, $A_k$), Transverse, Longitudnal and Average wave velocity (V$_t$, V$_l$ and V$_a$ in m/s), Debye Temperature ($\theta_D$ in K), Pugh's Ratio (k), Chen and Tian Vickers hardness ($H^{C_V}$, $H^{T_V}$ in GPa), Lame's first and second parameter ($\lambda$, $\mu$ in GPa) of XVH$_3$ (X=Li,K) perovskites under external stress of 5 and 10 GPa.

| Stress | 5 GPa | | 10 GPa | |
| --- | --- | --- | --- | --- |
| Parameters | LiVH$_3$ | KVH$_3$ | LiVH$_3$ | KVH$_3$ |
| C$_{11}$ | 185.995 | 113.395 | 215.638 | 36.460 |
| C$_{12}$ | 46.602 | 65.820 | 52.392 | 93.592 |
| C$_{11}$-C$_{12}$ | 139.393 | 163.246 | 15.174 | 57.132 |
| C$_{11}$+2C$_{12}$ | 279.199 | 225.037 | 320.424 | 223.645 |
| C$_{44}$ | 45.824 | 61.112 | 51.297 | 66.685 |
| $CP = C_{12}$-C$_{44}$ | 0.778 | 4.708 | 1.095 | 26.907 |
| B$_V$ | 93.066 | 75.012 | 106.807 | 74.548 |
| B$_R$ | 93.066 | 75.012 | 106.807 | 74.548 |
| B | 93.066 | 75.012 | 106.807 | 74.548 |
| G$_V$ | 55.373 | 48.182 | 63.427 | 28.585 |
| G$_R$ | 53.099 | 42.171 | 60.251 | 9.015 |
| G | 54.236 | 45.177 | 61.839 | 55.279 |
| E$_V$ | 138.626 | 119.056 | 158.840 | 25.376 |
| E$_R$ | 133.842 | 106.547 | 152.145 | 24.857 |
| E | 136.242 | 112.871 | 155.506 | 123.765 |
| $v_V$ | 0.252 | 0.235 | 0.252 | 0.330 |
| $v_R$ | 0.260 | 0.263 | 0.263 | 13.066 |
| $v$ | 0.256 | 0.249 | 0.257 | 1.430 |
| $\zeta$ | 0.462 | 0.821 | 0.452 | 11.541 |
| $A_u$ | 0.214 | 0.713 | 0.264 | 0.345 |
| $A_k$ | 0.094 | 0.298 | 0.115 | 0.326 |
| V$_t$ | 4565.635 | 3909.246 | 4769.651 | 3542.125 |
| V$_l$ | 7972.604 | 6763.947 | 8344.175 | 5645.832 |
| V$_a$ | 5072.309 | 4339.607 | 5299.799 | 4235.564 |
| $\theta_D$ | 762.309 | 590.859 | 808.202 | 435.234 |
| k | 1.616 | 1.660 | 1.627 | 1.634 |
| $H^{C_V}$ | 7.996 | 7.268 | 8.782 | 6.987 |
| $H^{T_V}$ | 8.415 | 7.675 | 9.165 | 7.324 |
| $\lambda$ | 56.909 | 44.894 | 65.581 | 43.934 |
| $\mu$ | 54.236 | 45.177 | 61.839 | 44.267 |

Ranganathan-Anisotropy Index ($A_u$) for cubic crystals is computed from the equation as

$$A_u = 5 \left[ \frac{G_v}{G_R} - 1 \right] \qquad (20)$$

Kleinman's parameter ($\zeta$) for our compositions is computed as per equation given as

$$\zeta = \frac{c_{11} + 8c_{12}}{7c_{11} + 2c_{12}} \qquad (21)$$

The value of $\theta_D$ is precisely determined for each volume V using the elastic constants by spherical average of the three components of sound velocity. It is provided as

$$\theta_D = \frac{\hbar}{k_B} \sqrt[3]{n * 6\pi^2 \sqrt{V}} \sqrt{\frac{B}{M} f_v} \qquad (22)$$

where $k_B$ is the Boltzmann constant, $\hbar$ is Planck's constant, and n is the number of atoms in the primitive cell with volume V unit. A function of the Poisson ratio $v$, denoted as $f_v$, and M, the compound's mass, correspond to V.

$$f_v = 3 \sqrt{\frac{3}{2 \left[ \frac{2(1+v)}{3(1-2v)} \right]^{\frac{3}{2}} + \left[ \frac{(1+v)}{3(1-v)} \right]^{\frac{3}{2}}}} \qquad (23)$$

Another important mechanical parameter is the degree of hardness of a material. Pugh used the formula $H_B = G\frac{b}{c}$, where b is the dislocation's Burger vector and c is a constant for all metals with the same structural composition, to create a link between the shear modulus (G) and the Brinell hardness ($H_B$) of pure metals. Teter discovered the semi-empirical link between Vicker's hardness H$_V$ and the rigidity modulus (G) as H$^{T_V}$ ∼0.151G. Additionally, Chen et al. provided a semi-empirical relationship between the shear modulus (G) and the squared Pugh's ratio ($k = \frac{B}{G}$), which determines Vicker's hardness.

$$H^{C_V} = 2(k^{-2}G)^{0.585} - 3 \quad (24)$$

and the same has been used to compute the values as listed in Table 2. The compositions' transverse ($V_t$) and longitudinal ($V_l$) elastic wave velocities are also calculated using the relations listed as

$$V_l = \sqrt{\frac{3B + 4G}{\rho}}; V_t = \sqrt{\frac{G}{\rho}} \quad (25)$$

where $\rho$ represents the mass density of the composition. All mechanical properties of these compositions have been computed using equations (8-25) and are listed in table 2. Bulk modulus measures volume change under hydrostatic pressure, while shear modulus measures resistance to shape change while maintaining volume. A material's stiffness is indicated by its Young's modulus, sometimes referred to as its modulus of elasticity, which is the ratio of tensile stress to tensile strain. In summary, a stiffer material exhibits a greater Young's modulus value. The bulk, shear, and Young's moduli of LiVH$_3$ are found to be higher compared to KVH$_3$ at higher pressure hence as the table demonstrates. $\frac{B}{G}$ values can be used to identify a compound's ductility or brittleness; a ratio of less than 1.75 indicates brittleness, whereas a ratio of more than 1.75 indicates ductility [58]. The examined compositions are feably brittle in character, according to the listed values of k shown in Table 2. Crystal anisotropy influences features such as phonon modes, plastic deformation, and crack behavior, and is crucial for many technological applications. Understanding the elastic anisotropy of materials is essential to understanding these effects. The degree of anisotropy in the bonding between atoms within distinct planes can be measured using shear anisotropic factors like Ranganathan and Kube Anisotropy Index ($A_u$, $A_k$). Three factors can influence a material's behavior in different crystal orientations. As reported in table 2 all these compositions do not have value equals to one which means that they possess anisotropic behavior.

### 3.8. Thermodynamic Properties

The Gibbs2 code [48] has been used to estimate the thermodynamical properties of XVH$_3$ (X = Li, K), including thermal expansion, specific heat, volume, $\theta_D$ and entropy, as well as how these properties change with temperature (T) and pressure (P). We looked at these characteristics in the range of 0–600 K and pressure ranges from 0 to 10 GPa. The heat capacity ($C_v$) provides information on lattice vibration and assesses molecular mobility at a constant volume. From Figure 9(a) shows the variation of ($C_v$) with temperature at constant pressure. According to Dulong Petit's law ($C_v$) rises with an increase in temperature and become saturated at elevated temperatures. The degree of a system's disorder is measured by its entropy (S). Figure 9(b) shows the variation of Entropy with temperature at constant pressure. There is an increase in entropy with rise in temperature. Also when the pressure is increased the degree of disorder reduces. The variation of thermal expansion coefficient ($\alpha$) varies with temperature and pressure, as shown in Fig. 9(c). Due to lattice compression, the pressure variation decreases with rising temperature, while the temperature variation shows the reverse tendency. The volume increases with respect to temperature but sharply decreases with respect to pressure, as shown in Fig. 10(a) which shows that these compositions are susceptible to both pressure and temperature. The elastic constant, heat capacity, melting point, and other solid-state physical properties are connected to the Debye temperature by the QHA Debye model [43]. Figure 10(b) shows the dependency of $\theta_D$ upon pressure and temperature. At 0K and 0 GPa the computed value of $\theta_D$ is 500 K. As the temperature rises and the external pressure remained constant, the $\theta_D$ dropped. Furthermore, we observed that the $\theta_D$ increases with increasing pressure at a constant temperature. The heat capacity at stable pressure (Cp) increases monotonically with temperature. Over a range of pressure values, this pattern holds consistency as shown in Figure 10(c).

## 4. Conclusion

Using the DFT-based WIEN2k code, the structural, electronic, mechanical thermoelectric and thermodynamic characteristics of the Vanadium-based perovskites XVH$_3$ (X = Li, K) are investigated. According to the structural investigations, all compositions in space group $Pm - 3m$ [no. 221] have a stable cubic structure. They are metallic in nature, as shown by plots of the energy band gap and the density of states. Notably, the FP-LAPW approach gives us a enthalpy of formation as -248.86 kJ/mol.H$_2$ and -457.27 kJ/mol.H$_2$ for XVH$_3$ (X = Li, K) perovskites, which indicates these perovskites are thermodynamically stable. Furthermore, the enhanced high temperature hydrogen storage cyclability of these compositions is indicated by the high decomposition temperatures. The compositions under investigation have promising hydrogen storage potential, as demonstrated by their respective gravimetric hydrogen storage capacities of 3.25%, and 4.97% for XVH$_3$ (X = Li, K). Furthermore, the mechanical stability of these materials is supported by the Born stability criterion. They also showed strong thermoelectric responses, which made them suitable materials for use in thermoelectric devices as alternative green energy sources in addition to their potential for storing hydrogen.

## 5. Declaration of Competing Interest

The authors have no conflict of interest for the work reported in this manuscript.

## CRediT authorship contribution statement

**Anupam:** Software, Workstation, Data generation. **Shyam Lal Gupta:** Conceptualization, Methodology, Data curation, Writing - original draft Writing - review and editing.

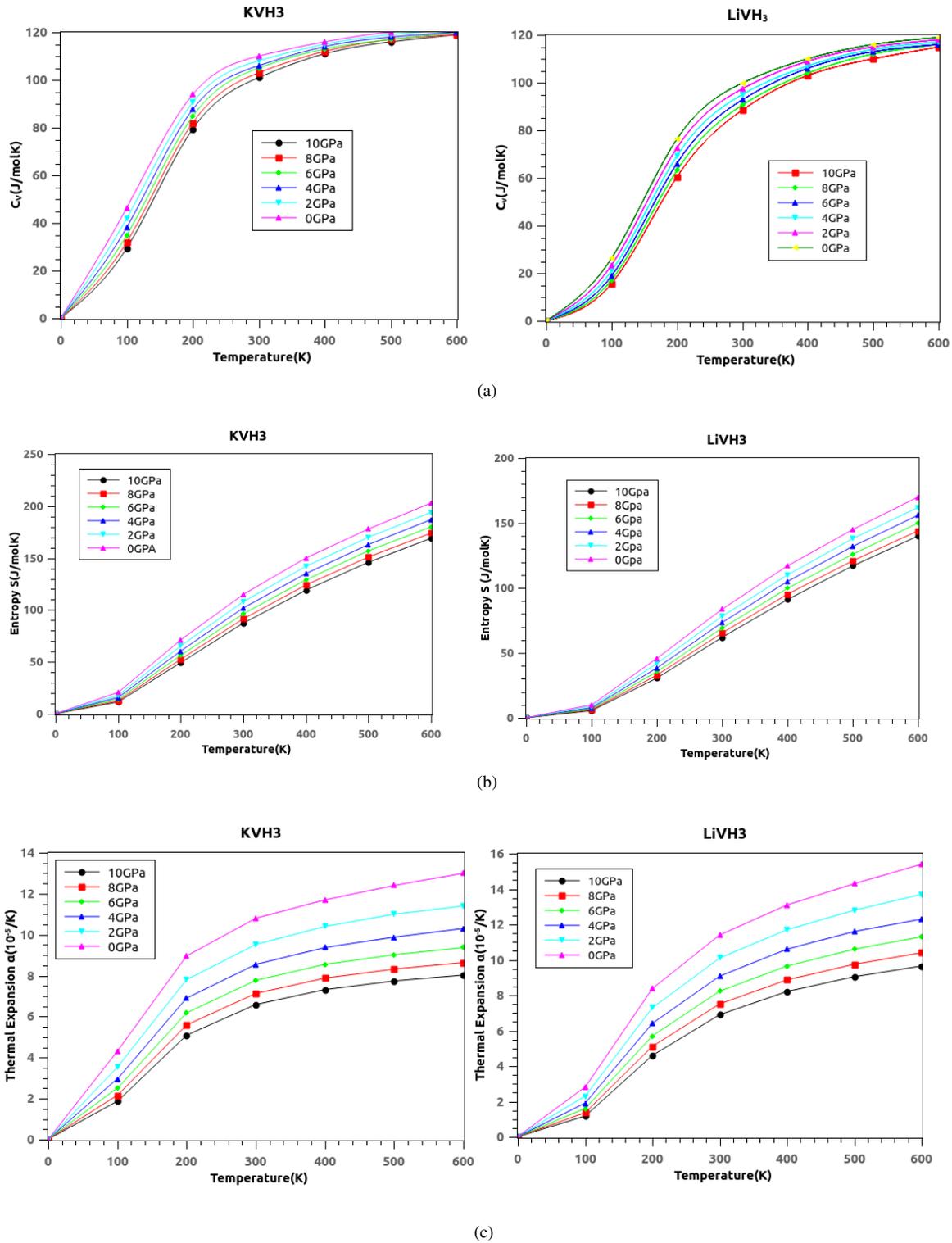

(a)

(b)

(c)

**Figure 9:** Variation of $C_v$, Entropy and Thermal expansion with temperature and pressure for $XVH_3$(X=Li,K) perovskites

**Sumit Kumar:** Software, Workstation, Data generation.
**Ashwani Kumar:** Software, Workstation, Data generation.
**Sanjay Panwar:** Software, Workstation, Data generation.
**Diwaker:** Conceptualization, Methodology, Data curation, Writing - original draft Writing - review and editing.

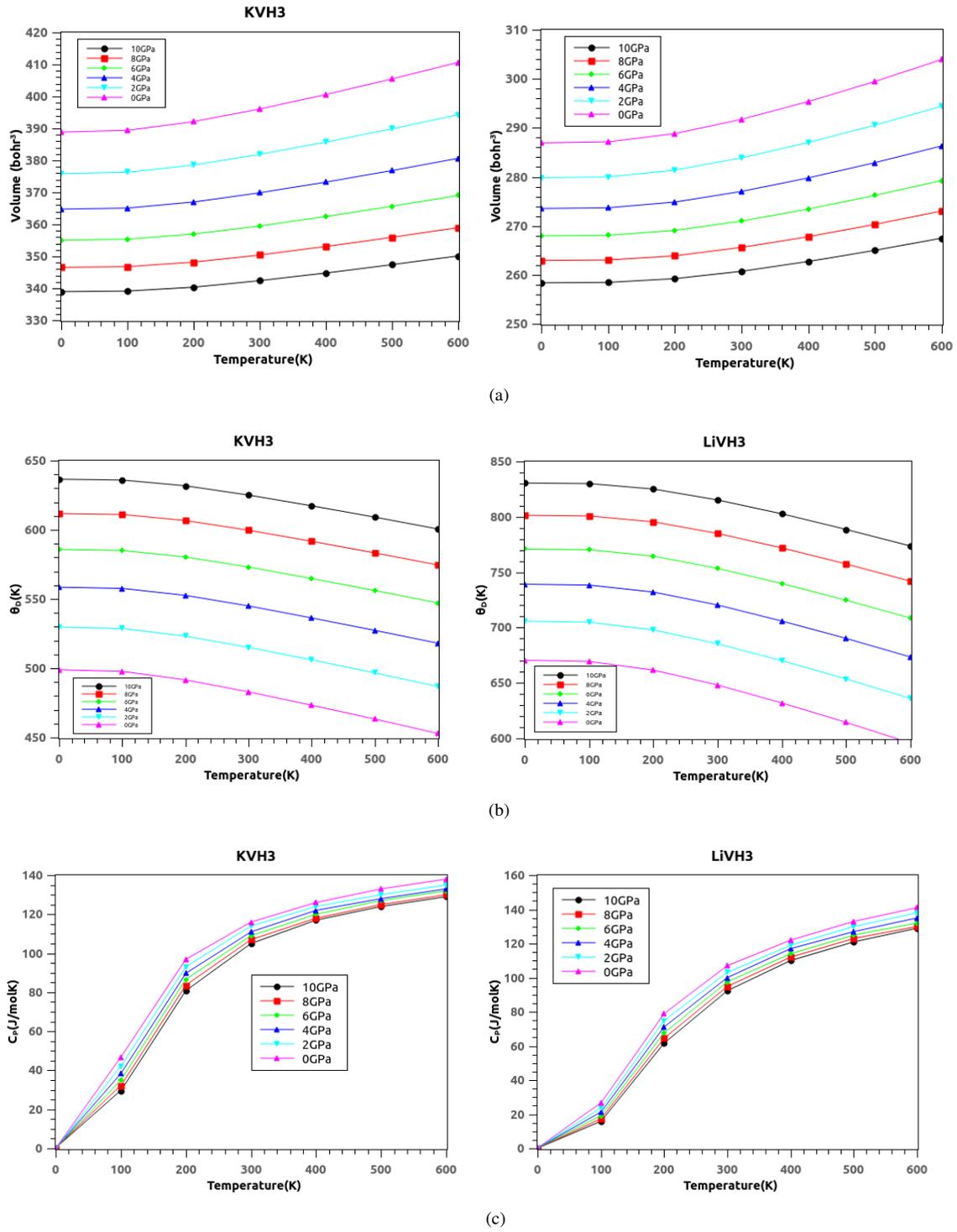

**Figure 10:** Variation of Volume, Debye temperature and $C_p$ with temperature and pressure for $XVH_3$(X=Li,K) perovskites